\documentclass{sig-alternate}

\usepackage{graphicx}
\usepackage{epsfig}
\usepackage{epstopdf}
\usepackage{algorithm} 
\usepackage{algorithmic}
\usepackage{url}
\usepackage{comment}
\usepackage{enumerate}

\begin{document}
\conferenceinfo{DOLAP'12,} {November 2, 2012, Maui, Hawaii, USA.} 
\CopyrightYear{2012} 
\crdata{978-1-4503-1721-4/12/11} 
\clubpenalty=10000 
\widowpenalty = 10000

\title{Benchmarking Summarizability Processing in XML Warehouses with Complex Hierarchies}

\numberofauthors{3}
\author{
	\alignauthor
	Chantola Kit
	\alignauthor
	Marouane Hachicha\\ \vspace{0.1cm}
	\affaddr{Universit{\'e} de Lyon (ERIC)}\\
	\affaddr{5 av. P. Mend{\`e}s-France}\\
	\affaddr{69676 Bron Cedex, France}\\
	\email{first.last@univ-lyon2.fr}
	\alignauthor
	J{\'e}r{\^o}me Darmont
}
     
\date{}

\maketitle

\begin{abstract}

Business Intelligence plays an important role in decision making. Based on data warehouses and Online Analytical Processing, a business intelligence tool can be used to analyze complex data. Still, summarizability issues in data warehouses cause ineffective analyses that may become critical problems to businesses. To settle this issue, many researchers have studied and proposed various solutions, both in relational and XML data warehouses. However, they find difficulty in evaluating the performance of their proposals since the available benchmarks lack complex hierarchies. In order to contribute to summarizability analysis, this paper proposes an extension to the XML warehouse benchmark (XWeB) with complex hierarchies. The benchmark enables us to generate XML data warehouses with scalable complex hierarchies as well as summarizability processing. 
We experimentally demonstrated that complex hierarchies can definitely be included into a benchmark dataset, and that our benchmark is able to compare two alternative approaches dealing with summarizability issues.

\end{abstract}




\section{Introduction}
\label{sec-introduction}

With the world in competitive business and innovation, Business Intelligence (BI) and Data Warehouses (DWs) play a major role in decision support. BI is famed for its complex analyses in many areas such as customer behavior, business trends, and new opportunities. BI heavily relies on DWs for storing and managing data. Without effective DWs, organizations cannot extract the required information for decision support. In addition, Online Analytical Processing (OLAP) is a notable BI tool for information analysis. To allow OLAP analyses, DWs are modeled as multidimensional structures~\cite{AgrawalGS97, RizziALT06}, where an observed fact is described by several hierarchical dimensions (star or snowflake schema).

DWs are designed to collect possible historical and concurrent data from various resources. This effort may produce complex hierarchies~\cite{BeyerCCOPX05, MalinowskiZ06, Riz07} that impact effective multidimensional analysis. For example, a book in an online bookstore may belong to more than one category (e.g., database and Web), and some book-sales may miss customer information if they are sold to anonymous customers. Such hierarchical information may lead to analysis errors, e.g., if we are trying to calculate the total quantity of books of computer science, which is the parent category of database and Web, or the total amount of sales by customer region. Such errors critically lead to poor or wrong decisions. These issues are termed as summarizability issues~\cite{HornerS05, HurtadoGM05, LenzS97, MansmannS07, PedersenJD99} and have been extensively surveyed by Maz\`on et al.~\cite{MazonLT09}. 

Concurrently, many people, companies, and organizations share information via online business, management, and social networking. The eXtensible Markup Language (XML) is widely used as a standard format for representing, transferring, and sharing data on the Web. Moreover, the XML tree-like format tends to be more efficient in representing complex hierarchical data than traditional relations of DWs~\cite{BeyerCCOPX05}.

Finally, many benchmarks support performance evaluation, including both relational and XML decision support benchmarks. Still, researchers have difficulty in evaluating the performance of summarizability processing algorithms, since there exists no benchmark with scalable complex hierarchies. In order to fill in this gap, we extend in this paper the XML data warehouse benchmark~\cite{XWeb} with scalable complex hierarchy generation and summarizability processing.

The remainder of this paper is organized as follows. Section~\ref{sec-background} formally defines complex hierarchies and discusses related work. Section~\ref{sec-xweb-ch} presents the data model and the algorithms we design for complex hierarchy generation, a query workload inducing summarizability issues and the performance metrics of our benchmark. Section~\ref{sec-experimental-demonstration}  illustrates the applicability of our benchmark through experiments. Section~\ref{sec-conclusion} gives the conclusion and perspectives of this work..

\newpage

\section{Background}
\label{sec-background}

In this section, we characterize complex hierarchies, which cause summarizability issues and present the research literatures related to decision support benchmarks. 

\subsection{Complex Hierarchies}
\label{subsec-complex-hierarchies}
We term a dimension hierarchy as complex if it is both non-strict and incomplete.

\subsubsection{Non-Strict Hierarchy}
\label{subsubsec-non-strict-hierarchy}

A hierarchy is non-strict \cite{AbelloSS06,MalinowskiZ06,Tor03} or multiple-arc \cite{Riz07} when an attribute is multivalued. In other terms, from a conceptual point of view, a hierarchy is non-strict if the relationship between two hierarchical levels is many-to-many instead of one-to-many. For example, in a dimension describing products, a product may belong to several categories instead of just one. 

Similarly, a many-to-many relationship between facts and dimension instances may exist \cite{Riz07}. For instance, in a sale data warehouse, a fact may be related to a combination of promotional offers rather than just one.  

\subsubsection{Incomplete Hierarchy}
\label{subsubsec-incomplete-hierarchy}

A hierarchy is incomplete \cite{MazonLT09}, non-covering \cite{AbelloSS06,MalinowskiZ06,Tor03} or ragged \cite{Riz07} if an attribute allows linking between two hierarchical levels by skipping one or more intermediary levels. For example, in a dimension describing stores, the $store/city/state/country$ hierarchy allows a store to be located in a given region without being related to a city (stores in rural areas). 

Similarly, facts may be described at heterogeneous granularity levels. For example, in our sale data warehouse, sale volume may be known at the store level in one part of the world (e.g., Europe), but only at a more aggregate level (e.g., country) in other geographical areas.

\subsubsection{Discussion}
\label{subsubsec-discussion}

Dimension hierarchy characterizations vary widely in the literature related to multidimensional models. For example, Beyer et al. name complex hierarchies ragged hierarchies \cite{BeyerCCOPX05}, while Rizzi defines ragged hierarchies as incomplete only \cite{Riz07}. Malinowski and Zim{\`{a}}nyi also use the terms of complex generalized hierarchy \cite{MalinowskiZ06}. Even though they include incomplete hierarchies, they do not include non-strict hierarchies. Thus, we prefer the term complex hierarchies.

Finally, note that some papers, addressing the summarizability problem, differentiate between intradimensional relationships and fact-to-dimension relationships \cite{MazonLT09}. By contrast, as Pedersen et al. \cite{PedersenJD99}, we consider that summarizability issues and solutions are the same in both cases, since facts may be viewed as the finest granularity in the dimension set.

\subsection{Related Work}
\label{subsec-related-work}

\subsubsection{Relational Decision Support Benchmarks}
\label{subsubsec-relational-decision-support-benchmarks}

The Transaction Processing Performance Council (TPC) defines standard benchmarks and publishes objectives and verifiable performance evaluations to the industry. The TPC currently supports two decision support benchmarks: TPC-H and TPC-DS. 

TPC-H's database follows a classical \textit{product-order-supplier} relational model~\cite{tpch}. Its workload is constituted of twenty-two SQL-92, parameterized, decision support queries, and two refreshing functions that insert tuples into and delete tuples from the database, respectively. Query parameters are randomly instantiated following a uniform law. Three primary metrics are used in TPC-H. They describe performance in terms of power, throughput, and a combination of these two criteria. Power and throughput are the geometric and arithmetic mean values of database size divided by workload execution time, respectively.

Although decision-oriented, TPC-H's database schema is not a typical star-like data warehouse schema. Moreover, its workload does not include any explicit OLAP query. The TPC-DS benchmark addresses this shortcoming~\cite{tpcds}. TPC-DS' schema represents the decision support functions of a retailer under the form of a constellation schema with several fact tables and shared dimensions. TPC-DS's workload is constituted of four classes of queries: reporting, ad-hoc decision support, interactive OLAP, and extraction queries. SQL-99 query templates help randomly generate a set of about five hundred queries, following non-uniform distributions. The warehouse maintenance process includes a full Extract, Transform, and Load (ETL) phase, and handles dimensions with respect to their nature (non-static dimensions scale up while static dimensions are updated). One primary metric is proposed in TPC-DS to take both query execution and the maintenance phase into account.


The Star Schema Benchmark (SSB) has been proposed as a simpler alternative to TPC-DS~\cite{ssb}. It is based on TPC-H's database remodeled as a star schema. It is basically designed around an order fact table merged from two TPC-H tables. More interestingly, SSB features a query workload that provides both functional and selectivity coverages.


In TPC-H, TPC-DS, and SSB, scaling is achieved through a scale factor \textit{SF} that defines data size (from 1 GB to 100 TB). Both database schema and workload are fixed. The number of generated queries in TPC-DS also directly depends on \textit{SF}. TPC standard benchmarks aim at comparing the performances of different systems in the same experimental conditions, and are intentionally not very tunable. By contrast, the Data Warehouse Engineering Benchmark (DWEB) helps generate various ad-hoc synthetic data warehouse (modeled as star, snowflake, or constellation schemata) and workloads that include typical OLAP queries~\cite{DarmontBB07}. DWEB targets data warehouse designers and allows testing the effectiveness of designed choices or optimization techniques in various experimental conditions thanks to complete set of parameters. Thus, it may be viewed more like a benchmark generator than a single benchmark. Nevertheless, DWEB's complete set of parameters makes it somewhat difficult to master.

\subsubsection{XML Decision Support Benchmarks}
\label{subsubsec-xml-decision-support-benchmarks}

There are many XML benchmarks such as the Michigan Benchmark~\cite{RunapongsaPJCK97}, MemBer~\cite{BeyerCCOPX05}, X-Mach~\cite{BöhmeR03}, XMark~\cite{SchmidtWKCMB02}, XOO7 ~\cite{BressanDLLLNW01} and XBench~\cite{Yao04xbenchbenchmark}. 
Unfortunately, none of these benchmarks exhibits any decision support feature. XWeB is the only decision support XML benchmark~\cite{XWeb}.

As in SSB, XWeB's DW schema is a simplified, snowflake version of TPC-H's schema. Moreover, XWeB's DW schema is logically and physically represented in XML. Since existing XML DW architectures mostly differ in the way dimensions are handled and the number of XML documents that are used to store facts and dimensions, XWeB exploits a unified model that is close to XCube~\cite{HummerBH03}. In this representation, an XML DW is composed of three XML documents at the physical level: \emph{dw-model.xml} defines the multidimensional structure of the warehouse (metadata); each $facts_f.xml$ document stores information related to set of facts $f$, including measure values and dimension references; and each $dimension_d.xml$ document stores dimension $d$'s hierarchy level instances.

Finally, XWeB's workload, expressed in XQuery, is constituted of twenty decision support queries labeled $Q01$ to $Q20$. The workload is structured in increasing order of complexity: reporting ($Q01$ to $Q03$), one dimension cubing ($Q04$ to $Q07$), two dimension cubing ($Q08$ to $Q11$), three dimension cubing ($Q12$ to $Q14$), and complex hierarchy querying ($Q15$ to $Q20$).

\subsubsection{Discussion}
\label{subsubsec-related-works-discussion}

XWeB includes only one complex hierarchy into its workload, i.e., \emph{part/category}. Complexity lies on the possible combination of $category$ instances in three levels, and queries are restricted by specific $part/category$ levels. Moreover, this complex hierarchy is not scalable and it does not cover all the cases of complex hierarchies defined in Section~\ref{subsec-complex-hierarchies}.

\section{Benchmark Specification}
\label{sec-xweb-ch}

We first present in this section our database model and complex hierarchies (Section~\ref{subsec-data-model}). We further describe how to generate complex hierarchies in the dataset (Section~\ref{subsec-generating-complex-hierarchies}). 
Then, we specify the benchmark's query workload, which operates onto complex hierarchies, induces summarizability issues (Section~\ref{subsec-query-workload}), and is aimed at being executed on the dataset to output performance metrics (Section~\ref{subsec-performance-metrics}).

\subsection{Data Model}
\label{subsec-data-model}

\begin{figure} [hbt]
\centering
\epsfig{file=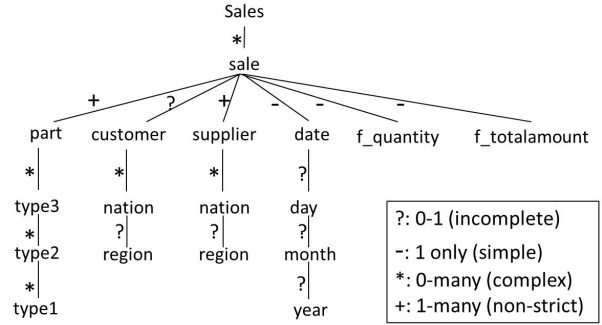, width=8cm}
\caption{Data model}
\label{fig-data-model}
\end{figure}

We model our benchmark's database after TPC-H's as illustrated in Figure~\ref{fig-data-model}. The data model consists of a $sales$ DW, a $sale$ fact, four dimensions: $part$, $customer$, $supplier$, and $date$, and two measures: $f\_quantity$ and $f\_totalamount$. Each dimension is subdivided into hierarchical levels. As specified by XWeB, the $part$ dimension contains three categorical levels that we label as type3, type2, and type1. Their instances are listed in Table~\ref{table-part-levels}. The $supplier$ and $customer$ dimensions possess two geographical levels: nation and region. The last dimension $date$ contains three levels: day, month, and year. In Figure~\ref{fig-data-model}, we borrow Annotated Tree Pattern's (APT)~\cite{PaparizosWLJ04} notations to specify the cardinality of relationships (edges); that is $?$: 0 or one, $-$: one only, $*$: 0 to many, and $+$: one to many. Note that relationships reveal complex hierarchies as defined in Section~\ref{subsec-complex-hierarchies}, i.e., incomplete and non-strict hierarchies. We can interpret the APT notations in Figure~\ref{fig-data-model} into complex hierarchies as follows: $?$: only incompleteness is possible when it is zero, $-$: incompleteness and non-strictness are impossible (simple hierarchy), $*$: incompleteness and non-strictness are possible, and $+$: only non-strictness is possible.

\vspace{-0.4cm}

\begin{table}[hbt]
\centering
\caption{Part hierarchical levels}
\label{table-part-levels}
{\tiny
\begin{tabular}{|l|l|} 
\hline
type3& ECONOMY, LARGE, STANDARD, PROMO, MEDIUM, SMALL\\ \hline
type2& ANODIZED, BURNISHED, BRUSHED, POLISHED, PLATED\\ \hline
type1& COPPER, NICKEL, STEEL, TIN, BRASS\\ \hline
\end{tabular}}
\end{table}

At the logical level, we utilize an instance of \emph{dw-model.xml} to represent the test DW (Figure~\ref{fig-logical-schema}). We exclude attribute values that store fact and dimension IDs from fact and dimension tags for brevity.

At the physical level, fact and dimension instances are stored in a set of XML documents, namely $facts_1.xml$ = $f\_sale.xml$, $dimension_1.xml$ = $d\_part.xml$, $dimension_2.xml$ = $d\_customer.xml$, $dimension_3.xml$ = $d\_supplier.xml$, and $dimension_4.xml$ = $d\_date.xml$. 

\begin{figure}[hbt]
\centering
{\scriptsize
\begin{tabbing}
<?xml version=`1.0' encoding=`UTF-8'?>\\
<dw-model>\\
<fac\=t id=`sale' path=`f\_sale.xml'>\\
\> <dim\=ension idref=`part' path=`d\_part.xml'>\\
\>\> <typ\=e3><type2><type1/></type2></type3>\\
\> </dimension>\\

\> <dimension idref=`customer' path=`d\_customer.xml'>\\
\>\> <nation><region/></nation>\\
\> </dimension>\\

\> <dimension idref=`supplier' path=`d\_supplier.xml'>\\
\>\> <nation><region/></nation>\\
\> </dimension>\\

\> <dimension idref=`date' path=`d\_date.xml'>\\
\>\> <day><month><year/></month></day>\\
\> </dimension>\\

\> <measure id=`f\_quantity'/>\\
\> <measure id=`f\_totalamount'/>\\

</fact>\\
</dw-model>
\end{tabbing}
}
\vspace{-0.4cm}
\caption{$dw$-$model.xml$}
\label{fig-logical-schema}
\end{figure}
\vspace{-0.4cm}

\subsubsection{Non-strict Hierarchies} 
According to the reality of sales, only the $part$ and $supplier$ dimensions may have non-strict characteristics w.r.t. $sale$ facts, since a sale may consist of many parts (special promotion or offer) and many suppliers (many suppliers or a supplied company whose branches are located in various countries). Moreover, non-strict hierarchies can appear among all hierarchical levels of the $part$ dimension, as a $part$ may belong to many categories and a finer level category may belong to many coarser level categories. The $customer$ and $supplier$ dimensions may also contain non-strict hierarchies at the nation level in case the $customer$ or $supplier$ is a company whose branches are located in various countries. However, the $nation/region$ hierarchy is strict since a nation belongs to one region only. Lastly, the $date$ dimension cannot be non-strict since a sale is restricted to a specific date.

\subsubsection{Incomplete Hierarchies}
Incomplete hierarchies can occur between the $sale$ and $customer$ dimensions in case of anonymous customer, or whose information including nation and region is unknown. Moreover, incompleteness may happen on hierarchical levels of the four dimensions  because of missing values.

\subsection{Generating Complex Hierarchies}
\label{subsec-generating-complex-hierarchies}

As we pointed out in Section~\ref{subsec-data-model}, complexity in hierarchies may occur at any level of any dimension. We divide the ``degree'' of complexity into four kinds: simple, incomplete only, non-strict only, and complex (both incomplete and non-strict). To generate a DW with scalable complex hierarchies, we propose some parameters: $fact\_number$, $incomplete\_per$ $centage$, $nonstrict\_$ $percentage$, and $nonstrict\_number$. 

The number of facts to be generated can be specified by $fact\_number$. The occurrence probability of incomplete or non-strict hierarchy instances among the total number of dimension instances are defined by $incomplete\_percentage$ or $nonstrict\_percentage$, respectively. Lastly,  the number of non-strict hierarchy instances in a dimension is specified by $nonstrict\_number$. Consequently, four kinds of hierarchies can be specified as follows. 

\begin{enumerate}
\scriptsize
\item{Simple (default): $incomplete\_percentage$ = 0 and $nonstrict\_$ $percentage$ = 0}
\item{Incomplete: $incomplete\_percentage$ > 0 and $nonstrict\_per$ $centage$ = 0}
\item{Non-strict: $nonstrict\_percentage$ > 0, $nonstrict\_number$ > 1 and $incomplete\_percentage$ = 0}
\item{Complex: $incomplete\_percentage$ >0, $nonstrict\_percentage$ > 0 and $nonstrict\_number$ > 1}
\end{enumerate}

\subsubsection{Simple Hierarchies}

\begin{figure}[hbt]
\begin{center}
\epsfig{file=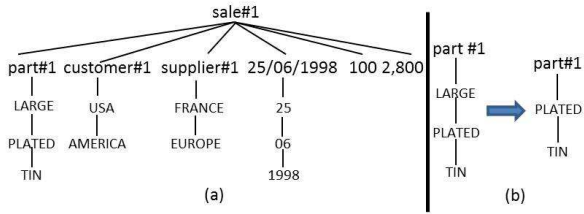, , width= 8 cm}
\end{center}
\caption{(a) Sale and (b) Incomplete hierarchy generation examples}
\label{fig-sale-incomplete-examples}
\end{figure}

XWeB directly provides simple hierarchies, so there is nothing to enhance at this point. Let us nonetheless provide a running example of fact, modeled as a data tree in Figure~\ref{fig-sale-incomplete-examples}(a), which we reuse in the following subsections. The example shows that on 25/06/1998, customer \#1 from USA of America region bought 100 parts (part \#1), costing 2,800 from supplier \#1, which is located in France, Europe. 

\subsubsection{Incomplete Hierarchies}
\label{subsubsec-incomplete-hierarchies}

Incomplete hierarchies are generated according to the $fact\_$ $number$ and $incomplete\_percentage$ parameters. For example, if $fact\_number$ = 10, 40 dimension instances (1 fact = 4 dimension instances) are generated. A dimension instance includes the dimension and its hierarchical instances. Then, if we set $incomplete\_percentage$ to 50, a hierarchy is randomly removed among every 2 (100/50) dimension instances to form an incomplete hierarchy. 

Algorithm~\ref{alg-incomplete-hierarchy} depicts incomplete hierarchy generation. An incomplete hierarchy is generated on the dimension specified by the input parameter $dim$. We use $ic\_check$ to verify the occurrences of incomplete hierarchies, that is at least one incomplete hierarchy exists in the given dimension and each level of hierarchy in the dimension is randomly chosen to be removed (for statement of the algorithm). Let us look at the example in Figure~\ref{fig-sale-incomplete-examples}(b). Suppose that dimension $part$ in Figure~\ref{fig-sale-incomplete-examples}(a) is randomly selected for incompleteness on its first level (``LARGE''). Then, the level ``LARGE'' is deleted from the $part$ dimension.

\begin{algorithm}[hbt]
\caption{: Incomplete hierarchy generation}
\label{alg-incomplete-hierarchy}
\begin{algorithmic}
\begin{small}
\STATE{\texttt{Input:$dim$//~target dimension}}
\STATE{$ic\_check$ = false}
\STATE{while $ic\_check$ is false $\{$}
\STATE{~~~~~for each level of $dim$ $\{$}
\STATE{~~~~~~~~~~randomly determine if current level bears incompleteness}
\STATE{~~~~~~~~if current level is selected $\{$}
\STATE{~~~~~~~~~~~~~remove current level from $dim$}
\STATE{~~~~~~~~~~~~~set $ic\_check$ to true}
\STATE{~~~~~~~~$\}$//~end if}
\STATE{~~~~~$\}$//~end for}
\STATE{$\}$//~end while}
\end{small}
\end{algorithmic}
\end{algorithm}

\subsubsection{Non-strict Hierarchies}
\label{subsubsec-non-strict-hieararchies}
We use the $nonstrict\_percentage$ parameter to specify the distribution of non-strictness. In addition, $nonstrict\_number$ is used to specify the maximum number of non-strict instances in a dimension. Note that at least two non-strict instances are generated (definition of non-strictness in Section~\ref{subsubsec-non-strict-hierarchy}). 

Algorithm~\ref{alg-non-strict-hierarchy} depicts non-strict hierarchy generation. Non-strict instances are formed in an array whose rows represent non-strict dimension instances and columns represent hierarchical levels. The $dim$ input parameter is used to specify the non-strict dimension instance. The algorithm randomly assigns $nonstrict\_number$ dimension instances to an $ns\_array$ (non-strict array). Finally, the $ns\_array$ is translated into an XML segment. Let us look at the example in the upper part of Figure~\ref{fig-complex-hierarchy-example}. Suppose that the $supplier$ dimension instance of Figure~\ref{fig-sale-incomplete-examples} (a) is chosen for non-strictness with $nonstrict\_number$ = 4. As a result, an array of four rows is created as in the left upper part of Figure~\ref{fig-complex-hierarchy-example}. The right upper part of the figure shows the data tree translated from the array. In this example, we can see that a sale is supplied by two suppliers (supplier\#1 and supplier\#2), and each of the suppliers owns two branches located in two different nations.

\begin{algorithm}[hbt]
\caption{: Non-strict hierarchy generation}
\label{alg-non-strict-hierarchy}
\begin{algorithmic}
\begin{small}
\STATE{\texttt{Input:$nonstrict\_number, dim$}}
\STATE{$ns\_array = null$}
\STATE{while $nonstrict\_number > 0$ \{ //~more dimension to be added}
\STATE{~~~~~randomly select dimension as $rand\_dim$}
\STATE{~~~~~add $rand\_dim$ to $ns\_array$}
\STATE{~~~~~$nonstrict\_number - 1 $}
\STATE{$\}$//~end while}
\STATE{$return$~$ns\_array$}
\end{small}
\end{algorithmic}
\end{algorithm}

\subsubsection{Complex Hierarchies}

\begin{figure} [hbt]
\centering
\epsfig{file=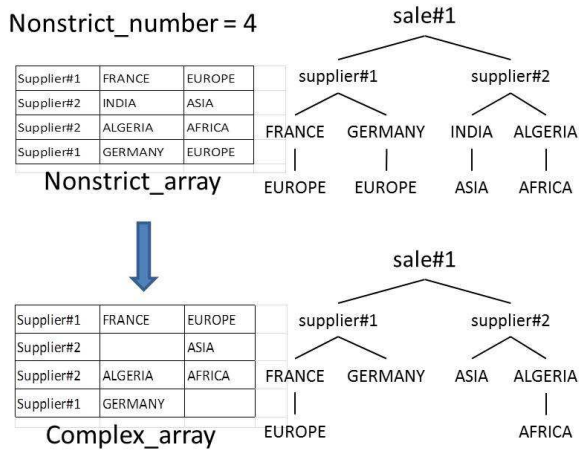, width=8cm}
\caption{Non-strict and complex hierarchies generations examples}
\label{fig-complex-hierarchy-example}
\end{figure}

Complex hierarchies occur when a dimension instance is chosen for both incompleteness and non-strictness. We use $nonstrict\_percentage$ (greater than zero) and $nonstrict\_number$ (greater than one) to specify non-strictness to be generated on the target dimension. Moreover, $incomplete\_percentage$ is set to be greater than zero to specify that incompleteness is also generated, especially over non-strict dimension instances.
Algorithm~\ref{alg-complex-hierarchy} depicts complex hierarchy generation. Firstly, the algorithm generates a non-strict array using Algorithm~\ref{alg-non-strict-hierarchy}. 
Then, the algorithm uses $ic\_check$ to confirm that at least one non-strict instance is randomly selected. Then incomplete hierarchies are generated only on randomly selected non-strict dimension instances using Algorithm~\ref{alg-incomplete-hierarchy}. The lower part of Figure~\ref{fig-complex-hierarchy-example} illustrates an example. Here, a $supplier$ dimension instance is chosen for both non-strictness (as in the upper part of the figure) and incompleteness. Consequently, an array (left lower part of Figure~\ref{fig-complex-hierarchy-example}) is generated from $ns\_array$ by Algorithm~\ref{alg-complex-hierarchy}. Finally, this array is translated into a data tree as in the right lower part of Figure~\ref{fig-complex-hierarchy-example}, where ``EUROPE'' and ``INDIA'' are deleted.

\begin{algorithm}[hbt]
\caption{: Complex hierarchy generation}
\label{alg-complex-hierarchy}
\begin{algorithmic}
\begin{small}
\STATE{\texttt{Input:$ns\_array$}}
\STATE{$ic\_check = $ false}
\STATE{while $ic\_check$ is false $\{$}
\STATE{~~~~~for each row of $ns\_array$ $\{$//dimension}
\STATE{~~~~~~~~~~randomly determine if dimension bears incompleteness}
\STATE{~~~~~~~~~~if current dimension is selected $\{$}
\STATE{~~~~~~~~~~~~~~$gen\_ic\_dim($current dimension$)$//~call to Algorithm~\ref{alg-incomplete-hierarchy}}
\STATE{~~~~~~~~~~~~~~set $ic\_chek$ to true}
\STATE{~~~~~~~~~~$\}$//~end if}
\STATE{~~~~~$\}$//~end for}
\STATE{$\}$//~end while}
\end{small}
\end{algorithmic}
\end{algorithm}

\subsection{Query Workload}
\label{subsec-query-workload}

Since complex hierarchy queries in XWeB run only on the $part$ dimension, we complement its workload to cover all possible complex hierarchies on all dimensions, as discussed in Section~\ref{subsec-data-model}. Our benmark's workload complements XWeB's to include up to 4-dimension cubing, and also supports both simple and complex hierarchies. 

The following list itemizes our workload's queries on complex hierarchies that build 4-dimension cubes (4D), basic aggregation operations, and some OLAP operations. Queries are presented in natural language for space constraints. 
A 4D cube is extracted by Q21, i.e., total quantity and amount of sales in groups of part, customer, supplier, and date dimensions. 
We can also extract a 4D cube at specific hierarchical level of dimensions as in Q22, i.e., min quantity of sales among the groups of dimensions:  customer at nation level, part at type3 level, supplier at nation level, and date at day level.
Moreover, we can slice the cube into a 3D cube with max aggregation by Q23, i.e., max of total amount among the groups of month, part's type2, supplier's nation, and customer's region. Finally, a 3D cube with average of total amount in groups of supplier's region, part's type1, customer's region, and year is also built by Q24.

\begin{itemize}
\item Q21: sum of $f\_quantity$, $f\_totalamount$ from $part$, $customer$, $supplier$, $date$ group by $part$, $customer$, $supplier$, $date$
\item Q22: min of $f\_quantity$ from $customer$, $part$, $supplier$, $date$ group by $nation$, $type3$, $nation$, $day$
\item Q23: max of $f\_totalamount$ from $date$, $part$, $supplier$, $customer$ group by $month$, $type2$, $nation$, $region$
\item Q24: average of $f\_totalamount$ from $supplier$, $part$, $customer$, $date$ group by $region$, $type1$, $region$, $year$
\end{itemize}

\subsection{Performance Metrics}
\label{subsec-performance-metrics}

In this benchmark, we define two performance metrics for summarizability processing algorithms. 

The first metric is quantitative: it is response time, i.e., the execution time of the query workload over a given dataset. Whether the overhead of summarizability processing can be distinguished from query processing or not depends on cases, but it is always included in the global execution time.

The second metric is qualitative: when running the benchmark, we check whether aggregation queries provided the expected results, i.e., we check whether summarizability issues are correctly handled. To do so, we check if the resulted groups are not duplicated, the total of aggregation values is equal to grand total, if average is the division of total and its number, min is the least value, or max is the highest value. 

\section{Experimental Demonstration}
\label{sec-experimental-demonstration}

To illustrate the feasibility and usefulness of our benchmark, we report in this section experiments in which we compare two methods for processing summarizability issues in DWs with the help of our benchmark. Note that our objective is to show that our benchmark provides useful insights regarding the behavior and performance of such approaches, and not so much to actually compare them.

\subsection{Studied Algorithms}
\label{subsec-studied-algorithms}

The first approach for addressing summarizability issues we test in this paper is a reference approach by Pedersen et al. \cite{PedersenJD99} (labeled \texttt{Pedersen} in the remainder of this section), which we adapt to the XML DW case. \texttt{Pedersen} transforms dimension and fact instances to enforce summarizability by using two algorithms named \texttt{MakeCovering} and \texttt{MakeStrict}. 
\texttt{MakeCovering} inserts new values, exploited from metadata and/or expert advice, into the missing hierarchical levels to ensure that mappings to coarser hierarchical levels are covering/complete. 
\texttt{MakeStrict} avoids ``double counting'' by ``fusing'' multiple values in a parent hierarchical level into one ``fused'' value, and then linking the child value to the fused value. Fused values are inserted into a new hierarchical level in-between the child and the parent. Consequently, reusing this new level for computing coarser level aggregate values leads to correct aggregation results. \texttt{MakeCovering} and \texttt{MakeStrict} transform both the DW schema and data, and are applied once in a static way.

The second approach we test is a new, dynamic approach called Query-Based Summarizability \cite{Marouane12} (labeled \texttt{QBS} in the remainder of this section). \texttt{QBS} deals with summarizability issues by firstly avoiding multiplying the aggregation of measure instances of a hierarchical level when rolling up to a coarser level in non-strict hierarchies. Thus, when building the set of groups with respect to a grouping criterion, multiple values in the coarser level are fused into one single "fused value". Secondly, when rolling from a hierarchical level up to a coarser level, measure instances of the finer level that are not present in the coarser level must still be agregated (incomplete hierarchies). Thus, when building a group, all "missing instances" are grouped into an artificial "Other" group. By contrast to \texttt{Pedersen}, \texttt{QBS} does not transform schema nor data and applies automatically, on the fly, at query time.

\subsection{Experimental Configuration}
\label{subsec-experimenal-configuration}

\subsubsection{System Configuration}
\label{subsubsec-system-configuration}
Our experiments are done on a Toshiba laptop with an Intel(R) Core(TM) i7-2670QM CPU @ 2.20GHz, 4.00 GB of memory, and 64-bit Windows 7 Home Premium Service Pack 1. The algorithms are implemented in Java JDK 1.7 using the SAX parser to read XML data.

\subsubsection{Experimental Setup}
\label{subsec-experimental-setup}

We use DWs with complex hierarchies scaling in size from 50,000 to 250,000 facts as in the first row of Table~\ref{table-data-size}. The second row ranges simple hierarchy data in kilobytes (27~MB minimum and 134~MB maximum). The third, fourth, and fifth rows list the size of DWs with 5\% incomplete, non-strict, and complex hierarchies respectively. The sixth, seventh, and eighth rows denote the size of DWs with 50\% incomplete, non-strict, and complex hierarchies respectively.

\begin{table}[hbt]
\centering
\caption{Dataset size (KB)}
\label{table-data-size}
{\scriptsize
\begin{tabular}{|l|r|r|r|r|r|r|r|r|} 
\hline
No. Facts&50,000&100,000&150,000&200,000&250,000\\ 
\hline
Simple&27,700&55,390&82,800&110,577&138,015\\ \hline
Incomplete 5\%&27,626&55,242&82,543&110,249&137,573\\ \hline
Non-strict 5\%&28,669&57,328&85,671&114,422&142,786\\ \hline
Complex 5\%&28,376&56,742&84,791&113,252&141,319\\ \hline
Incomplete 50\%&25,020&50,030&74,769&99,842&124,601\\ \hline
Non-strict 50\%&35,412&70,826&105,914&141,397&176,527\\ \hline
Complex 50\%&32,522&65,031&97,263&129,839&162,088\\ \hline
\end{tabular}}
\end{table}

Among the workload of queries, we focus on four queries with various number of dimensions (labeled $n$: 1D to 4D), and select the most detailed hierarchy levels for grouping since they form more complex groups (Table~\ref{table-group-by-dimensions}). We roll up the queries to levels \emph{day}, \emph{type3}, \emph{nation}, and \emph{nation} of the \emph{date}, \emph{part}, \emph{customer}, and \emph{supplier} dimensions, respectively. 

\begin{table}[hbt]
\centering
\caption{Group by $n$-dimension queries}
\label{table-group-by-dimensions}
{\scriptsize
\begin{tabular}{|l|l|l|l|l|} 
\hline
$n$&part&customer&supplier&date\\ 
\hline
1D&&&&day\\ \hline
2D&type3&&&day\\ \hline
3D&type3&nation&&day\\ \hline
4D&type3&nation&nation&day\\ \hline
\end{tabular}}
\end{table}

\subsection{Experimental Results}
\label{subsec-experimental-results}

The following subsections present our experimental results of comparing \texttt{QBS} and \texttt{Pedersen}. For \texttt{Pedersen}, we differentiate  between query execution time and preprocessing overhead, while it is impossible for \texttt{QBS}, as overhead is embedded within query execution.  

The following results focus on the response time metric, because both our implementations of \texttt{Pedersen} and \texttt{QBS} compute correct aggregates (the quality metric is met in both cases).

\subsubsection{Results on Simple Hierarchies}
\label{subsubsec-experimental-result-on-noncomplex-XWeb}

Our first comparison is run on simple hierarchies only and the results are shown in the left-hand side of Figure~\ref{fig-experimental-results}. Figure~\ref{fig-experimental-results}(a) shows that \texttt{QBS}' time performance increases linearly with data size and the number of dimensions in the query, except the 3D query on 50,000 facts, which incidentally bears lower grouping complexity. Figure~\ref{fig-experimental-results}(b) shows that the time performance of both approaches increases linearly w.r.t. data size and the number of dimensions used in queries. Moreover, \texttt{QBS} expectingly performs a little worse than \texttt{Pedersen} without overhead, but tends to perform a little better when \texttt{Pedersen}'s overhead is accounted for. 

However, both \texttt{QBS} and \texttt{Pedersen} consume a lot of time, especially when running the 4D query (about an hour). To find out the cause, we perform two more experiments, disassociating complex hierarchy processing time from group matching time. This is possible because  XWeB's data are originally summarizable. Figure~\ref{fig-experimental-results}(c) shows that enforcing summarizability in \texttt{QBS} does not much affect time performance, while group matching has a great impact that increases with the number of dimensions. Figure~\ref{fig-experimental-results}(d) confirms that \texttt{Pedersen} also spends most of its time processing group matching, while overhead consumes little time. We notice that, when processing group matching, we indeed need to check whether the group exists. Thus, we must check every hierarchy level instance in the whole group, which contains several instances from all dimensions. Doing so is very time consuming comparing to traditional aggregation, which only checks for the existing group as a whole. However, no approach dealing with summarizability can avoid this issue.

\subsubsection{Results on Complex Hierarchies}
\label{subsubsec-experimental-result-on-generated-complex-tpch}

\begin{figure*}[hbt]
\begin{center}$
\begin{array}{ccc}
\epsfig{file=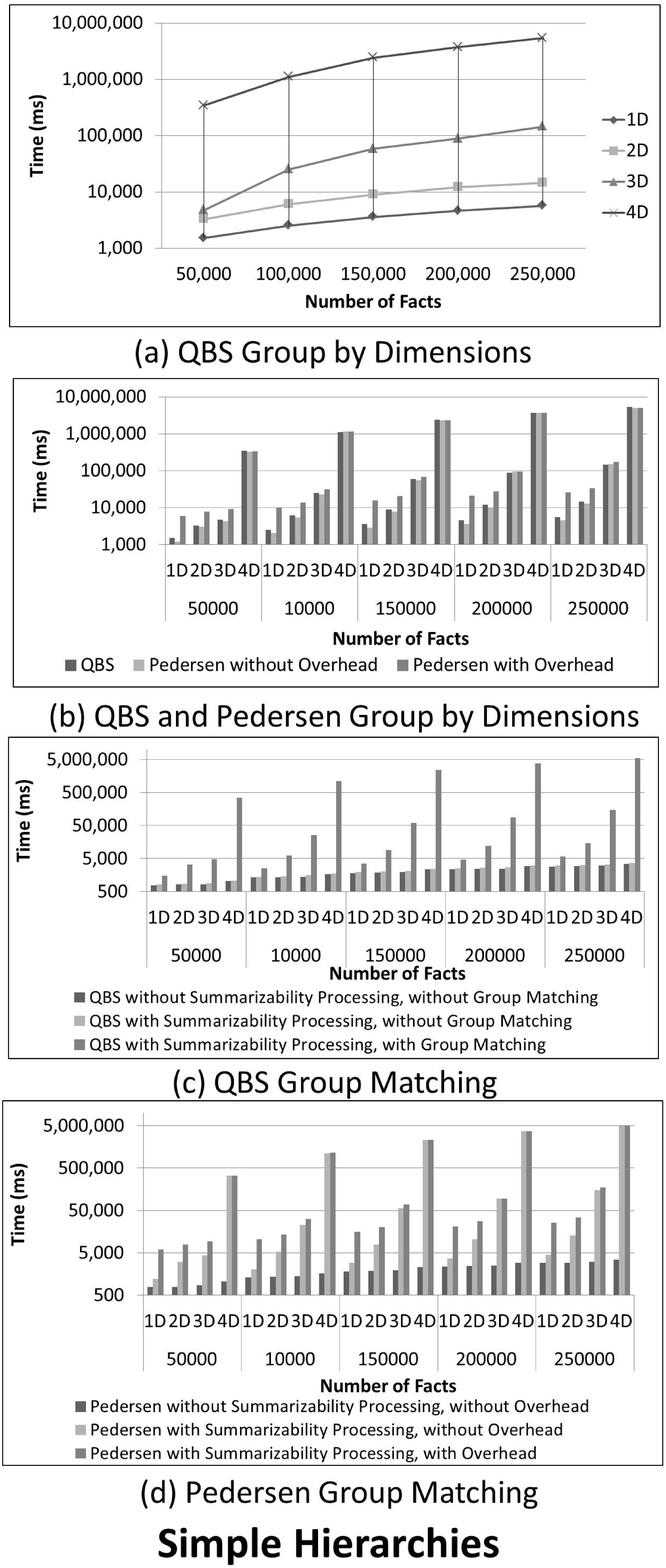, , width= 6 cm} &
\epsfig{file=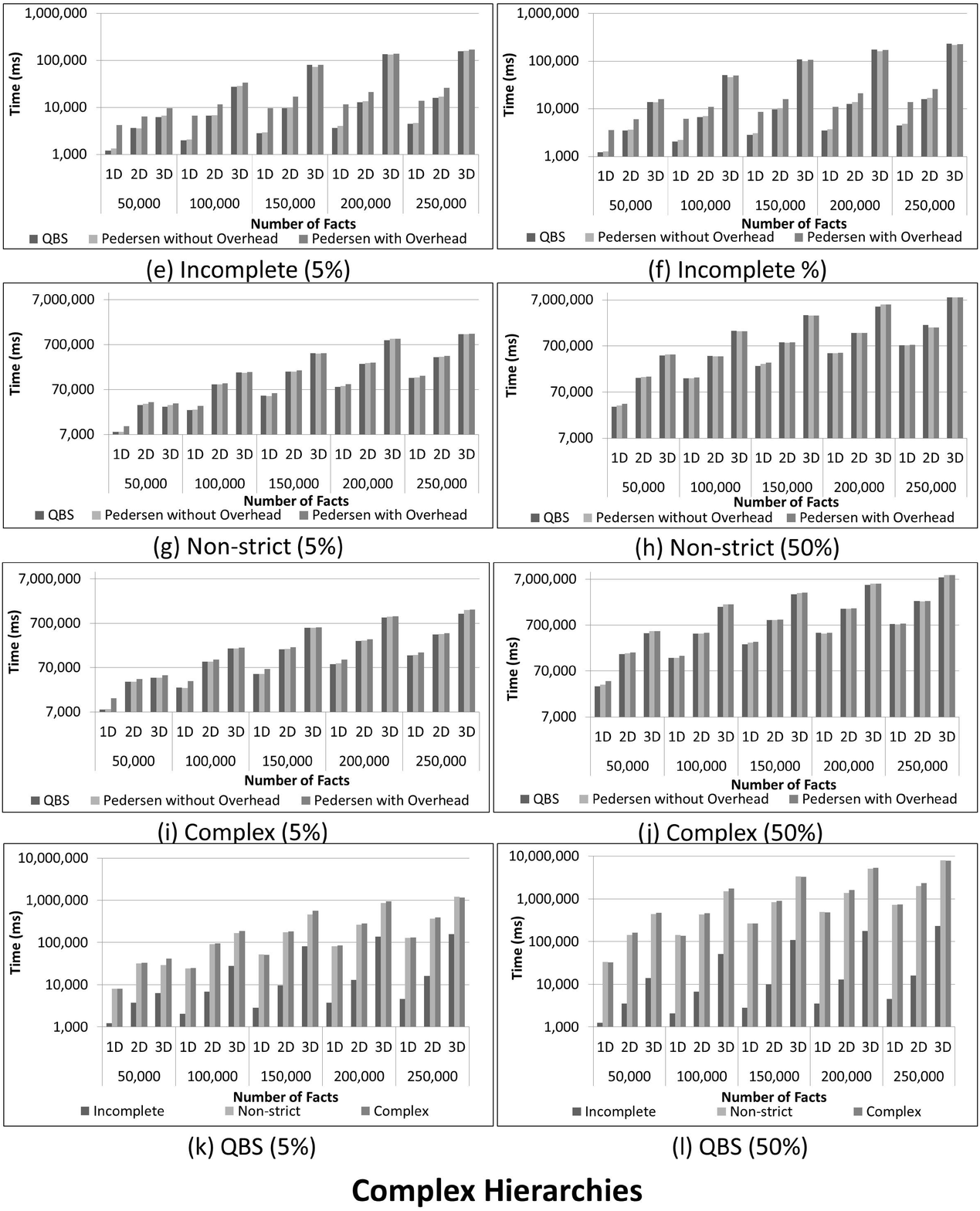, , width= 11.5 cm}
\end{array}$
\end{center}
\caption{Experimental results}
\label{fig-experimental-results}
\end{figure*}

Due to space limitations, we only present here our experiments on 5\% and 50\% incomplete, non-strict and complex hierarchies (the approximate minimum and maximum scales), but we did go through the whole range. The results we obtain are shown on the right-hand side of Figure~\ref{fig-experimental-results}.

\paragraph{Incomplete Hierarchies}

The results from Figures~\ref{fig-experimental-results}(e) and \ref{fig-experimental-results}(f) reveal two cases. When the number of dimensions is small (up to query 2D), \texttt{QBS} is comparable to \texttt{Pedersen} when overhead is excluded, and tends to perform better than \texttt{Pedersen} when overhead is included. For a larger number of dimensions (query 3D), both approaches are comparable. Both approaches actually have different tradeoffs. \texttt{QBS} takes less time when reading incomplete data, but more time to solve incompleteness, while the reverse is true for \texttt{Pedersen}. Thus, when the number of dimensions increases, the QBS's processing of incomplete hierarchies at query time is a handicap that evens global performances w.r.t. \texttt{Pedersen}. Still, we observe that both approaches are affected by the poor performance of group matching, which explains why we did not include query 4D in these experiments.

\paragraph{Non-Strict Hierarchies}
\label{sec:NonStrictHierarchies}

The results from Figures~\ref{fig-experimental-results}(g) and \ref{fig-experimental-results}(h) show similar trends to those of Figures~\ref{fig-experimental-results}(e) and \ref{fig-experimental-results}(f), because the tradeoffs in \texttt{QBS} and \texttt{Pedersen} are essentially the same for non-strictness management. However, non-strictness processing takes much more time than incompleteness processing in checking the existing non-strict instances in each dimension, as shown in Figures~\ref{fig-experimental-results}(k) and \ref{fig-experimental-results}(l) for \texttt{QBS}. Ultimately, we can again record that \texttt{QBS} is comparable to \texttt{Pedersen} without overhead, and a little better when overhead is included. 

\paragraph{Complex Hierarchies}
\label{sec:ComplexHierarchies}

The results from Figures~\ref{fig-experimental-results}(i) and \ref{fig-experimental-results}(j) bear similar results to the non-strict case, again because the cost of non-strictness processing is much higher than that of incompleteness processing (Figures~\ref{fig-experimental-results}(k) and \ref{fig-experimental-results}(l)). Group matching is indeed mainly impacted by non-strict hierarchies. However, in some cases, such as in the 3D query on 250,000 facts in Figure~\ref{fig-experimental-results}(k), \texttt{QBS} performs better in the complex case than in the non-strict case, because non-strict processing incidentally produces fewer complex groups, thus simplifying group matching.

\section{Conclusion}
\label{sec-conclusion}

To the best of our knowledge, our benchmark is the first XML data warehouse benchmark with complex hierarchies. It has been designed to conform to Gray's criteria (relevance, portability, scalability, and simplicity)~\cite{DBLP:books/mk/Gray93}. Our benchmark is \textit{relevant} since it refers to the TPC-H standard, while adding complex hierarchies that answer to precise engineering needs, i.e., summarizability processing performance testing. The benchmark is \emph{portable} as it is written in Java, making it easy to implement and connect to various systems, including XML DBMSs. It is \textit{scalable} by number of facts and complex hierarchy distributions. Finally, it is \textit{simple} since its model, which is inherited from XWeB, is a simplified, star-like version of TPC-H's.

Morevover, we demonstrate the use of our benchmark by comparing two approaches that address sumarizability issues when processing complex hierarchies, namely \texttt{Pedersen} and \texttt{QBS}. Our benchmark highlights two main insights. First, run-time summarizability management is feasible, since \texttt{QBS} performs almost as well as \texttt{Pedersen} dynamically, and would retain the same perfomance if DW shema or data changed, while \texttt{Pedersen} would have to be run again. Second, we show that both algorithms spend most of their time processing group matching. This is thus the main process to be optimized in future research on summarizability processing in XML environments.

Finally, a raw, preliminary version of our benchmark\footnote{\url{http://eric.univ-lyon2.fr/~ckit/DOLAP12.zip}} is freely available online as a NetBeans project\footnote{\url{http://netbeans.org}} . A more streamlined version is in the pipe and will be distributed under a Creative Commons license\footnote{\url{http://creativecommons.org/licenses/by-nc-sa/2.5/}}.

\newpage

In the future, we intend to integrate our benchmark with XWeB, including an XQuery parser that supports where clauses and OLAP operators (slice, dice, rotate, roll-up, drill-down, and cube). Furthermore, it would be interesting to add more unstructured business information (i.e., document-oriented XML data) such as in XMark and XBench into our benchmark.

\bibliographystyle{abbrv}
\bibliography{dolap20-kit}

\end{document}